\documentclass[aps,twocolumn,showpacs]{revtex4}
\usepackage{graphicx}
\usepackage[all]{xy}
\usepackage{amsmath}
\usepackage{amssymb}
\newcommand{\be}{\begin{equation}}
\newcommand{\ee}{\end{equation}}
\newcommand{\ben}{\begin{eqnarray}}
\newcommand{\een}{\end{eqnarray}}
\newcommand{\bes}{\begin{subequations}}
\newcommand{\ees}{\end{subequations}}

\newcommand{\bb}{\bibitem}
\begin{document}
\title{New models for two real scalar fields and their kinklike solutions}
\author{A. Alonso-Izquierdo,$^a$ D. Bazeia,$^{b,c,d}$  L. Losano,$^{c,d}$ and J. Mateos Guilarte$^a$}
\affiliation{
{\small{$^a$Departamento de Matematica Aplicada and IUFFyM   Universidad de Salamanca, Spain}\\
{$^b$Instituto de F\'\i sica   Universidade de S\~ao Paulo 05314-970 S\~ao Paulo SP, Brazil}\\
{$^c$Departamento de F\'\i sica   Universidade Federal da Para\'\i ba  58051-970 Jo\~ao Pessoa PB, Brazil}\\
{$^d$Departamento de F\'\i sica   Universidade Federal de Campina Grande 58109-970 Campina Grande PB, Brazil}\\
}}

\date{\today}
\begin{abstract}
{In this work we study the presence of kinks in models described by two real scalar fields in bi-dimensional space-time. We generate new two-field models,
constructed from distinct but important one-field models, and we solve them with techniques that we introduce in the current work. We illustrate the results with several
examples of current interest to high energy physics.}
\end{abstract}


\maketitle

\section{Introduction}

The presence of kinks and solitons in models described by real scalar fields is of direct interest to high energy physics \cite{V,MS} and other areas of nonlinear science \cite{WH,1}. To mention specific studies, in high energy physics kinks appear in very interesting systems introduced, for instance, in \cite{DG,DG2}. In condensed matter one can investigate domain walls in magnetic systems \cite{DW,DW2}, and nonlinear excitations in Bose-Einstein condensates \cite{KB,KB2}, to quote just a few examples.

In this work we focus on one-field and two-field models, in $(1,1)$ spacetime dimensions.
Two very interesting models described by a single real scalar field are known as the sine-Gordon and the $\phi^4$ models, engendering spontaneous symmetry breaking. The $\phi^4$ model is described by a fourth-order polynomial potential and supports kinklike solutions, whereas the sine-Gordon model is characterized by a non-polynomial potential and supports not only solitons but also multi-soliton and breather solutions. Fluctuations around the solitons and $\lambda \phi^4$ kinks, however, are governed by the $\ell=1$ and $\ell=2$ reflectionless Hamiltonians of a general family known from supersymmetric quantum mechanics \cite{Casahorran}. Moreover, a rich family of non-polynomial models with spontaneous symmetry breaking was proposed in \cite{Bordag}. The main feature of the family of kinks arising in this family is that the Hamiltonians governing the kink small fluctuations cover many of the remaining transparent SUSY Hamiltonians, see also \cite{Alonso}.

We start with these one-field models, which are described by polynomial and non-polynomial $W=W(\phi)$, and we then move on to two-field models constructed from the previous ones. Our aim is to identify kink solutions in these new models, which in general is a very difficult endeavor, as Rajaraman \cite{Ra} notices: \textit{This already brings us to the stage where no general methods are available for obtaining all localized static solutions (kinks), given the field equations. However, some solutions, but by no means all, can be obtained for a class of such Lagrangians using a little trial and error}. In this work we develop a technique which generates two-component kink solutions for two-field models in a straight-forward way avoiding the use of the trial and error method mentioned by Rajaraman. We mention, however, that there exist two scalar field theory models \cite{Aai1} and even models of three scalar field \cite{Aai2} such that all the kink solutions can be found due to the complete integrability of the analogue mechanical problem.

For simplicity, we use natural units, and then we redefine fields and coordinates such that fields and space and time are all dimensionless. The study starts in Sec.~\ref{sec:2}, and the one-field and two-field models are then used in Sec.~\ref{sec:3} to generate new models, described by two fields. In this section we deal with polynomial potential, and so, to enlarge the scope of the work, in Sec.~\ref{sec:4} we introduce another family of models, containing a non-polynomial function of the field $\chi$. We end the work in Sec.~\ref{sec:5}, where we introduce some comments and conclusions.

\section{Generalities}
\label{sec:2}

Let us first consider one-field models. We take the Lagrange density in the form
\be
{\cal L}=\frac{1}{2}\,\partial^\mu{\phi}\,\partial_\mu\phi-V(\phi)\,.
\ee
Here we deal with topological solutions, so we write the potential $V(\phi)$ in the form
\be
V(\phi)=\frac12 W_\phi^2
\ee
where $W=W(\phi)$, and $W_\phi$ stands for the derivative with respect to $\phi$, that is, $W_\phi=dW/d\phi$. The equation of motion for static field configuration is given by
\be\label{em1}
\phi^{\prime\prime}=\frac{dU}{d\phi}=W_\phi W_{\phi\phi}.
\ee
Here we are using $\phi^{\prime}=d\phi/dx$, etc. The energy density for static solutions can be written as
\be
\epsilon(x)=\frac12 \phi^{\prime\, 2}+\frac12 W_\phi^2= \frac{1}{2}(\phi^{\prime}-W_\phi)^2+\frac{dW}{dx}
\ee
for smooth superpotentials. We note that the energy is minimized to the value
\be
E_{\rm BPS}=|W(\phi(\infty))-W(\phi(-\infty))|
\ee
for field configurations that obey the first order equation
\be\label{first1}
\phi^\prime=W_\phi.
\ee
This is the Bogomol'nyi bound, and we can easily see that solutions to Eq.~{\eqref{first1}} also solve the equation of motion \eqref{em1}. The field configurations that solve the first-order equation are named Bogomol'nyi-Prasad-Sommerfeld (BPS) states \cite{BPS,BPS2}.
We note that since the potential does not see the sign of $W$, there are in fact two first-order equations, one for $W$, and the other with $W$ changed to $-W$. This is related with the spatial reflection symmetry $x\rightarrow -x$, which provides us with the kink/antikink solutions.

Two important models in the above class of models are the $\phi^4$ model,
where
\be
W(\phi)=\phi-\frac13\phi^3
\ee
and the sine-Gordon model, where
\be
W(\phi)=\sin(\phi) \,\,.
\ee
The potentials are, respectively,
\be
V(\phi)=\frac12(1-\phi^2)^2,
\ee
and
\be
V(\phi)=\frac12 \cos^2(\phi).
\ee
These models have solutions in the form, for the $\phi^4$ model,
\be
\phi(x)=\tanh(x)
\ee
and for the sine-Gordon model,
\be
\phi(x)={\rm arcsin}(\tanh(x))+k\pi, \;\;\; k=0,\pm1,\pm2,...,
\ee
where $k$ identifies one among the infinity of topological sectors of the sine-Gordon model.

Let us now consider two-field models. We start with the Lagrange density
\be
{\cal L}=\frac{1}{2}\,\partial^\mu{\phi}\,\partial_\mu\phi+\frac{1}{2}\,\partial^\mu{\chi}\,\partial_\mu\chi-V(\phi,\chi)\,.
\ee
For static configurations, the equations of motion become
\be
\phi^{\prime\prime}=\frac{\partial V}{\partial \phi}\qquad\text{and}\qquad\chi^{\prime\prime}=\frac{\partial V}{\partial \chi}\,.
\ee
We suppose that the potential $V(\phi,\chi)$ is given in terms of the superpotential $W(\phi,\chi)$, by
\be\label{vsuper}
V(\phi,\chi)=\frac12 {W_\phi^2}+\frac12{W_\chi^2},
\ee
where $W_\phi={\partial W}/{\partial \phi}$ and $W_\chi={\partial W}/{\partial \chi}$. Notice that the critical points of the superpotential $W(\phi,\chi)$ provide us with the set of vacua ${\cal M}=\{(\phi,\chi)\in \mathbb{R}^2:V(\phi,\chi)=0\}$ for the field theory model. The energy density has the form
\ben
\epsilon(x)&=&\frac{1}{2} \left( \phi^{\prime\,2}+\chi^{\prime\,2}+W_\phi^2+W_\chi^2\right)\, = \\
&=& \frac{1}{2} \left[ (\phi^{\prime}-W_\phi)^2+(\chi^{\prime}-W_\chi)^2 \right] +  dW \,\,. \nonumber
\een
The minimum energy solutions comply with
\be \label{orb1}
\phi^\prime=W_\phi \hspace{0.5cm} \mbox{and} \hspace{0.5cm} \chi^\prime= W_\chi\,
\ee
leading us to the BPS energy
\be\label{ebps}
E_{\rm BPS}=\left|W(\phi(\infty),\chi(\infty))-W(\phi(-\infty),\chi(-\infty))\right|
\ee
for smooth superpotentials. In terms of the superpotential, the equations of motion for static fields are written as
\ben
\phi^{\prime\prime}&=&W_{\phi}W_{\phi\phi}+W_{\chi}W_{\chi\phi}, \\
\chi^{\prime\prime}&=&W_{\phi}W_{\phi\chi}+W_{\chi}W_{\chi\chi},
\een
which are solved by the first order equations \eqref{orb1}, for $W_{\phi\chi}=W_{\chi\phi}$,
as we require in this work.
Solutions to these first order equations are BPS states, which solve the equations of motion. The sectors where the
potential has BPS states are named BPS sectors.

As an example, let us consider the model characterized by the superpotential
\be\label{mo}
W(\phi,\chi)=\phi-\frac13{\phi^3}-r\,\phi\,\chi^2\,
\ee
which has been studied by Shifman and Voloshin in the context of ${\cal N}=1$ supersymmetric Wess-Zumino models with two chiral superfields \cite{S,S2}. In the purely bosonic framework the presence of domain walls and its stability has been analyzed in the references \cite{B,B2,B3}, while in \cite{M,M2} the complete structure of this type of solutions is given in two critical values of the coupling between the two scalar fields by exploiting the integrability of the analogue mechanical system associated with this model. This well known model will be used in the following sections to illustrate the applicability of the novel procedure introduced in this paper which allows the identification of kinklike solutions new field theories.

The first-order ODE \eqref{orb1} in this case are written as
\be\label{wp1}
\phi^\prime=1-\phi^2-r\,\chi^2\hspace{0.5cm},\hspace{0.5cm}
\chi^\prime=-2r\,\phi\,\chi\, .
\ee
The potential is given by
\be
V=\frac12(1-\phi^2-r\chi^2)^2+2r^2\phi^2\chi^2\,,
\ee
which can be seen as an extension of the $\phi^4$ model to the case of two fields. Here we consider $r$ real and positive. The vacuum set comprises four elements:
\be
{\cal M}= \left\{ v_{1,2}=(\pm1,0),\,\, v_{3,4}=  \left(0,\pm\sqrt{1/r}\right)\right\} \,.
\ee
Associated with the superpotential \eqref{mo} we can find five BPS sectors (here we do not distinguish between kinks and antikinks) by analyzing the first order ODE \eqref{wp1}. Indeed this model is a very special case because an integrating factor can be calculated for the orbit equation extracted from \eqref{wp1}. The kink trajectories are given by
\be\label{orbits}
\phi^2+ \frac{r}{1-2r} \chi^2-\gamma \chi^\frac{1}{r}=1\hspace{0.3cm} \mbox{where} \hspace{0.3cm} \gamma \in (-\infty,\gamma_C]
\ee
with $\gamma_C=2r^{\frac{1}{2r}+1}/(1-2r)$ and $r\neq \frac{1}{2}$. For the range $\gamma \in (-\infty,\gamma_C)$ the formula \eqref{orbits} describes kinks connecting the vacua $v_3$ and $v_4$ with $E_{\rm BPS}=4/3$ while the value $\gamma=\gamma_C$ yields kinks linking the points $v_{1,2}$ with $v_{3,4}$ with $E_{\rm BPS}=2/3$. From construction all the kinks living in a specific topological sector are energy degenerate.

In particular the $\gamma=-\infty$ and $\gamma=0$ members of \eqref{orbits} correspond respectively to the one-component kink
\be\label{sol1}
\phi(x)=\pm\tanh(x)  \,\,\,\, , \,\,\,\, \chi(x)=0\, ,
\ee
and the two-component kink
\be\label{sol2}
\phi(x)=\pm\tanh(2\,r\,x) \,\, , \,\,\chi(x)=\pm\,\sqrt{\frac{1}{r}-2}\, \,\text{sech} (2\,r\,x)\,,
\ee
lying in the topological sector connecting the minima $v_3$ and $v_4$, see Figure 1.

\begin{figure}[h]
\includegraphics[height=3cm]{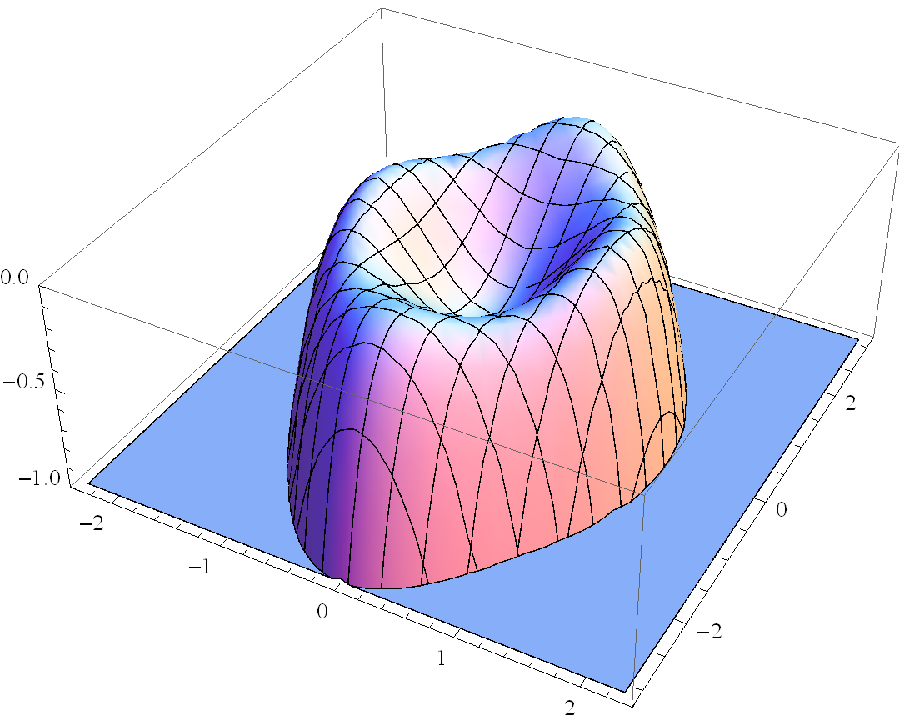} \hspace{0.5cm}
\includegraphics[height=3cm]{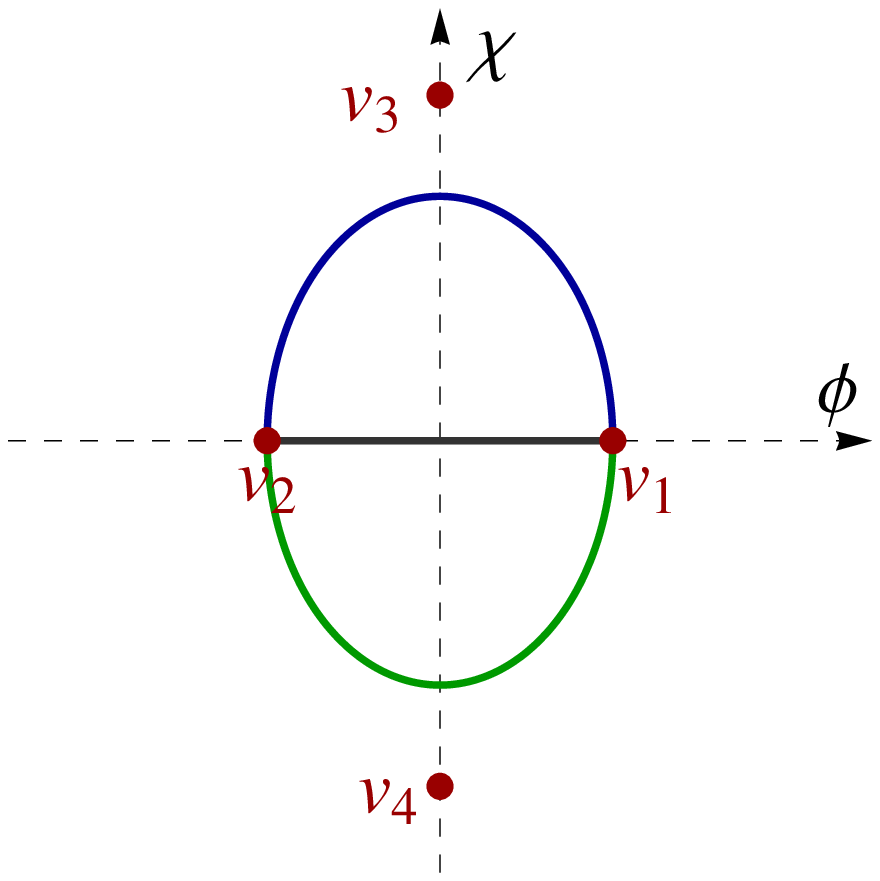}
\caption{Graphics of the potential $V(\phi,\chi)$ with $r=1/4$ and orbits of the kinks \eqref{sol1} and \eqref{sol2} in the internal plane.}
\end{figure}

As an illustrative sample we introduce a kink solution
\be\label{kink44}
\phi=\pm\frac{1}{2}( 1 \pm \tanh(x)) \hspace{0.3cm},\hspace{0.3cm} \chi=\pm\frac{1}{2}( 1\mp \tanh(x))
\ee
lying in the topological sectors joining the points $v_{3,4}$ with the points $v_{1,2}$ for the case $r=1$, see Figure 2.

\begin{figure}[h]
\includegraphics[height=3cm]{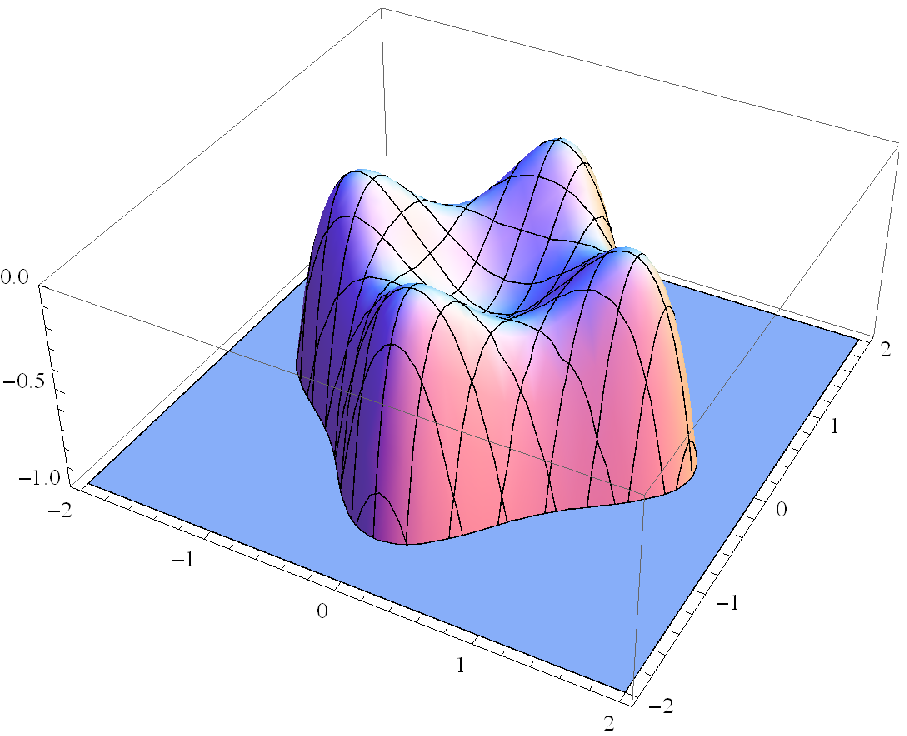} \hspace{0.5cm}
\includegraphics[height=3cm]{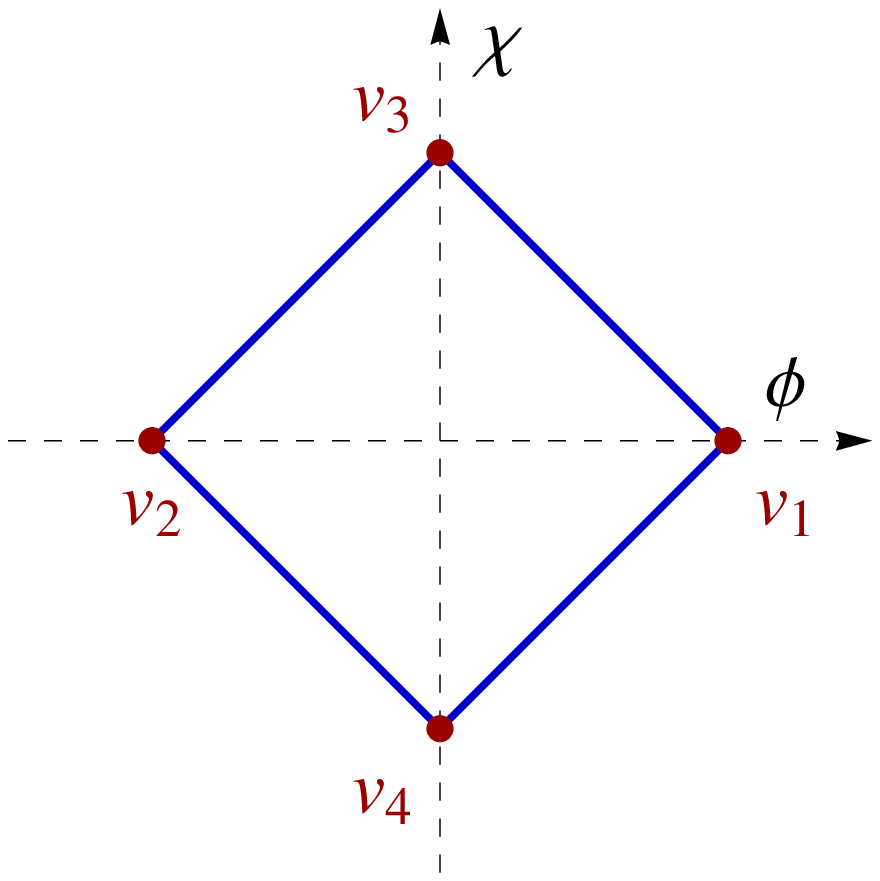}
\caption{Graphics of the potential $V(\phi,\chi)$ with $r=1$ and orbits of the kinks \eqref{kink44} in the internal plane.}
\end{figure}

We remark that different superpotentials can generate the same potential. Indeed in this model other superpotentials than \eqref{mo} have been identified for several particular values of the coupling constant $r$. This fact provides us with new degenerated BPS and non-BPS solutions in the topological sector joining the points $v_3$ and $v_4$, see \cite{M}.

\section{New models}
\label{sec:3}

To generate new two-field models and the accompanying static solutions, we proceed as follows: we start from the one-field model,
with the super potential written in the form
\be
W(\phi)=\int^{\phi} f(y)\, dy \,\, .
\ee
This gives the first-order equation
\be
\phi^{\prime}=f(\phi),
\ee
and for the $\phi^4$ and sine-Gordon models we have $f(\phi)=1-\phi^2$ and $f(\phi)=\cos(\phi)$, respectively.

Now, to introduce two-field models, we get inspiration on the previous model, given by Eq.~\eqref{mo} and we propose the following superpotential
\be\label{choice1}
W(\phi,\chi)=\int^\phi{f(y)dy}-r\,\phi\,\chi^2\,.
\ee
This generates the field potential term
\be\label{potentialchoice}
V(\phi,\chi)=\frac{1}{2} \left[f(\phi)-r \chi^2\right]^2+2r^2 \phi^2\chi^2 \,\, .
\ee
The critical points of the superpotential, determined by $W_\phi=f(\phi)-r\chi^2=0$ and $W_\chi=-2r\phi \chi=0$, provide us with the vacua of the model:
\be\label{vacua}
{\cal M}=\left\{ (\phi,\chi)\in \mathbb{R}^2 : (\phi^{(i)},0)\,\, , \,\, \Big(0,\pm \sqrt{\frac{f(0)}{r}} \Big) \right\}
\ee
where $\phi^{(i)}$, $i=1,\dots,n$ are the roots of $f(\phi)$. Therefore this kind of models involves $n+2$ vacua assuming that $f(0)\neq 0$. The static solutions are obtained from the first-order equations
\be\label{gwp1}
\phi^{\prime}=f(\phi)-r\,\chi^2 \hspace{0.3cm},\hspace{0.4cm}
\chi^{\prime}=-2r\,\phi\,\chi\,.
\ee
We can manipulate these equations to get
\be\label{ddp1}
\phi^{\prime\prime}+\left(4r\phi-\frac{df}{d\phi}\right)\phi^{\prime}-4r\phi\,f(\phi)=0\,.
\ee
Choosing $\phi(x)$ as the static solution of a one-field model, such that
\be\label{u1u2}
\phi^\prime=U(\phi)\,\,\,\, , \,\,\,\,\,\phi^{\prime\prime}=\frac{dU(\phi)}{d\phi}\,U(\phi)\,\,\,,
\ee
we can rewrite the above Eq.\eqref{ddp1} as
\be\label{uddp}
\frac{df}{d\phi}+P(\phi)f(\phi)-Q(\phi)=0\,,
\ee
where
\be\label{pq1}
P(\phi)=\frac{4r\phi}{U};\,\,\,\,\,\,Q(\phi)=4r\phi+\frac{dU}{d\phi}\,.
\ee
The function $U(\phi)$ is a particular solution of the linear ODE \eqref{uddp}, so we can write the general solution in the form
\be\label{fp}
f(\phi)= U(\phi) + C e^{-\int{P(\phi)\,d\phi}}
\ee
with $C$ being an integration constant. Plugging \eqref{fp} into \eqref{potentialchoice} we get a one-parameter family of potentials which have a non-trivial two-component kink solution, whose orbit
\be\label{orbita}
r\,\chi^2=f(\phi)-U(\phi)
\ee
emerges from the use of \eqref{gwp1}, \eqref{u1u2} and \eqref{fp}. Notice that strictly speaking the ODE \eqref{uddp} must be verified only on the kink orbit, in the rest of the internal plane a natural extension of the potential is considered.

We note that the first-order equations \eqref{gwp1} supports the orbit
$\chi(x)=0$, providing us with a second kink solution for our model. In this case, the static solutions $\phi(x)$, connecting neighbor minima located at the $\phi$-axis, is obtained from
\be\label{orbita0}
\frac{d\phi}{dx}=f(\phi)\,.
\ee
The expression $f(\phi)$ coincides with $U(\phi)$ only if the integration constant $C=0$. The general expression of $f(\phi)$ gives rise to a family of two-component field theory models, which admit a two-component kink solution whose first component coincides with the kink associated to $U(\phi)$.

The key step of the procedure appears in Eq.~\eqref{u1u2}, and it is inspired from Ref.~\cite{bd}. It works nicely for a variety of choices of $U(\phi)$, and the corresponding models include polynomial and non-polynomial functions.

\subsection{A first example}

To illustrate the above procedure with concrete examples, let us start considering
\be\label{u11}
U(\phi)=\kappa(a-\phi)(b+\phi)\,,
\ee
where $a,b,\kappa$ are real parameters and we obviously assume that $a,b\neq 0$ and $a\neq -b$. We use equation \eqref{fp} together with \eqref{u1u2} and \eqref{pq1} to obtain
\be\label{fp1}
f(\phi)=\kappa(a-\phi)(b+\phi)+C (a-\phi)^{n_1} (b+\phi)^{n_2}\,,
\ee
where
\be\label{n33}
n_1=\frac{4ra}{\kappa(a+b)};\,\,\;\,\,\,\, n_2=\frac{4rb}{\kappa(a+b)}\,.
\ee
This leads to the field potential term
\ben
V&=&\frac{1}{2} \Big[ \kappa (a-\phi)(b+\phi)+C(a-\phi)^{n_1}(b+\phi)^{n_2} - \nonumber\\
&& \hspace{0.4cm} - r \chi^2\Big]^2+ 2 r^2 \phi^2 \chi^2 \, . \label{potential11}
\een
Here we have the static solution extracted from \eqref{u1u2} for our choice of $U(\phi)$ in \eqref{u11}; it reads
\be\label{phi1}
\phi(x)=\frac{a-b}{2}\pm \frac{a+b}{2} \tanh\left(\frac{\kappa(a+b)x}{2}\right)\,,
\ee
and, from \eqref{orbita} and \eqref{fp1} we obtain
\be\label{chi1}
\chi(x)=\pm\sqrt{\frac{C}{r}} \, \left(a- \phi(x)\right)^{\frac{n_1}{2}}\, \left(b+\phi(x)\right)^{\frac{n_2}{2}}\,.
\ee
Dilatations and translations in the internal space allow us to relocate two vacua placed in the $\phi$-axis at the points $(\pm 1,0)$, such that without loss of generality we can assume that $a=1$ and $b=1$. If we restrict ourselves to potentials \eqref{potential11} with a quartic algebraic expression in the fields $\phi$ and $\chi$ we must impose the conditions $\frac{4ar}{(a+b)\kappa}=1$ and $\frac{4br}{(a+b)\kappa}=1$, or equivalently $\kappa=2r$. In this case we get the family of potentials
\be\label{potential22}
V=\frac{1}{2} [(2r+C)(1-\phi^2)-r \chi^2]^2 + 2 r^2 \phi^2 \chi^2
\ee
where ${\cal M}=\{ v_{1,2}=(\pm 1,0), v_{3,4}=(0,\pm \sqrt{2+C/r})\}$ comprises four elements provided that $C>-2r$.
The two-component kinks
\be \label{kink22}
\phi(x)=\tanh (2r x) \hspace{0.3cm},\hspace{0.3cm} \chi(x)=\pm\sqrt{\frac{C}{r}} \,{\rm sech}\,(2rx)
\ee
whose kink orbit is given by $r \chi^2=C(1-\phi^2)$ connect the points $v_{3,4}$ and $E_{\rm BPS}=\frac{4}{3}(C+2r)$. The expression \eqref{potential22} can be written as
\be \label{potential33}
V=(2r+C)^2 \left[ \frac{1}{2} ( 1- \phi^2- \overline{r}\chi^2 )^2  +2 \overline{r}^2 \phi^2 \chi^2 \right]
\ee
where $\overline{r}=r/(2r+C)$. A re-parametrization of the spatial variable $\overline{x}=(2r+C)x$ allows us to identify the present example with the potential introduced in the previous section. In this sense if we choose $C=0$ we get the one-component topological kink solutions \eqref{sol1}. For any other choice of the constant $C$ the solution \eqref{kink22} plays the role of the two-component kink \eqref{sol2}. The comparison is straightforward when the constant $2r+C$ in \eqref{potential22} is unity, see Figure 1. This works as a test for the procedure introduced in this work in a well-known two-field theory model.

If we consider the special case $b=0$ in \eqref{u11}, such that $U=\kappa \phi(a-\phi)$, the use of \eqref{fp} leads to the function
\be
f(\phi)=\kappa \phi (a-\phi) + C (a-\phi)^\frac{4r}{\kappa}
\ee
which generates the field potential
\be
V=\frac{1}{2} [ \kappa \phi(a-\phi)+C(a-\phi)^\frac{4r}{\kappa} -r\chi^2]^2 + 2r^2 \phi^2\chi^2
\ee
Because we are interested in quartic potentials in this section we set $\kappa=2r$. The previous formula becomes
\be
V=\frac{1}{2} \left[ (a-\phi)(a C +(2r-C)\phi) -r\chi^2\right]^2 + 2r^2 \phi^2\chi^2\,,
\ee
whose zeroes are located at ${\cal M}= \{ v_1=(a,0), v_2=(-\frac{aC}{2r-C},0),v_{3,4}=(0,\pm a \sqrt{\frac{C}{r}})\}$. The equation \eqref{orbita} leads to the kink orbits $\chi = \pm \sqrt{C/r} \,(a-\phi)$ which connect the points $v_{3,4}$ with $v_1$, see Figure 3. The kink solutions are
\be \label{kink55}
\phi=\frac{a}{2}(1+\tanh (arx)) \hspace{0.2cm},\hspace{0.2cm}
\chi=\pm \frac{a}{2} \sqrt{\frac{C}{r}}(1-\tanh (arx))
\ee
whose energy is $E_{\rm BPS}=a^3(C+r)/3$.

\begin{figure}[h]
\includegraphics[height=3cm]{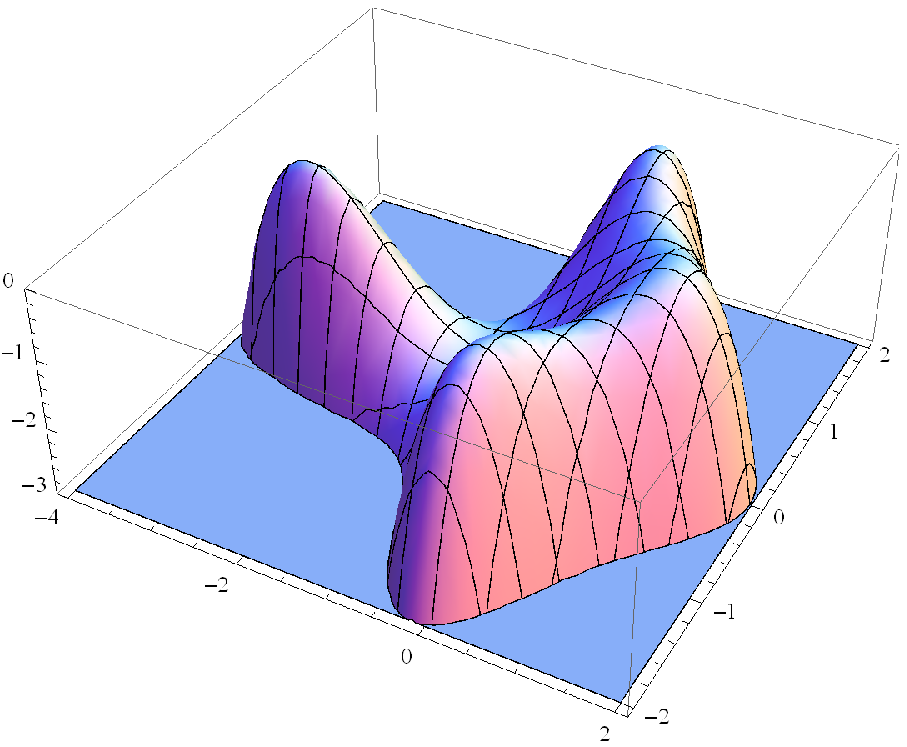} \hspace{0.5cm}
\includegraphics[height=3cm]{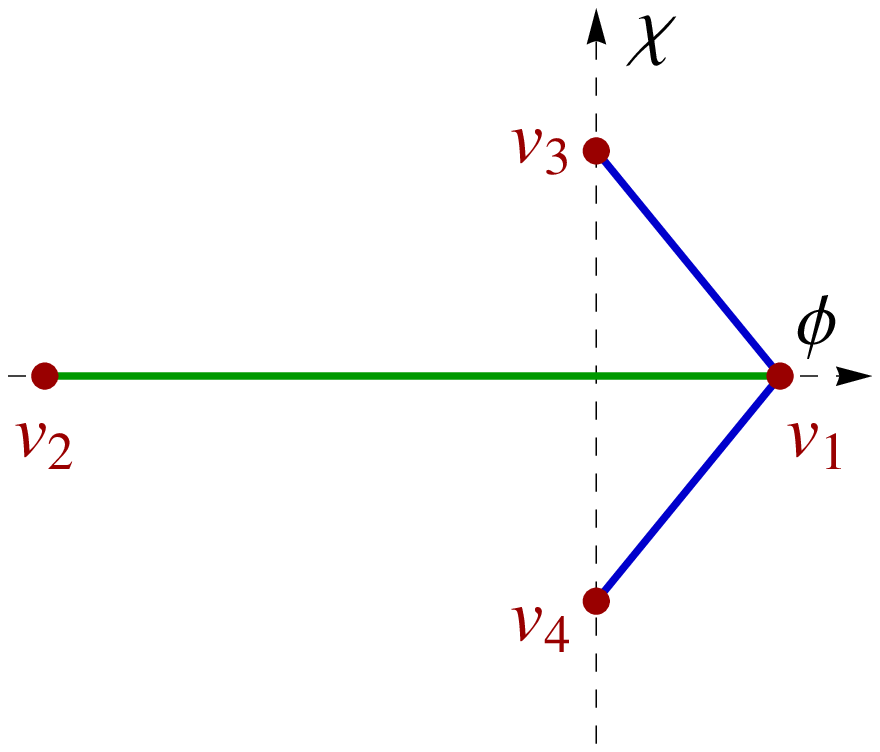}
\caption{Graphics of the potential $V(\phi,\chi)$ with $r=a=1$, $C=3/2$ and orbits of the kinks \eqref{kink55} and \eqref{kink56} in the internal plane.}
\end{figure}

As previously mentioned it is easy to identify the one-component kink linking the vacua $v_{1,2}$ for this model. We have
\be \label{kink56}
\phi=a + \frac{2ar}{C-2r-e^{2arx}} \hspace{0.3cm},\hspace{0.3cm} \chi=0 \,\, ,
\ee
whose energy is $E_{\rm BPS}=\frac{4a^3r^3}{3(C-2r)^2}$.

Notice that if we consider $r=a=C=\kappa/2=1$ in \eqref{kink55} we recover the solutions \eqref{kink44} of the test model introduced in the previous section.

The above illustration shows that the procedure works nicely. Thus, below we introduce new families of models using
adequate choices of the parameters.

\subsubsection{A family of models}

In the previous section we restrict ourselves to quartic potentials. Here we shall introduce expressions with higher degree. Let us choose $a=1$, $b=1$, $\kappa=2r/n$ in \eqref{fp1} and \eqref{n33} with $n$ a positive integer. Besides we redefine the coupling constant $r$ as $r=n\overline{r}/2$. Here we have
\be\label{fp11}
f(\phi)=(1-\phi^2)\left( \overline{r}+C(1-\phi^2)^{n-1}\right)\,,
\ee
which determines the potentials
\be
V= \frac{1}{2} \Big[(1-\phi^2)\left( \overline{r}+C(1-\phi^2)^{n-1}\right)-\frac{n\overline{r}}{2} \chi^2\Big]^2 + \frac{n^2 \overline{r}^2}{2} \phi^2\chi^2
\ee
These potentials involve two distinct behaviors depending on $n$ being odd or even.

In the case for $n$ even, there are six degenerate minima at the points ${\cal M}_e=\{v_{1,2}=(0,\pm\sqrt{2(\overline{r}+C)/(n\overline{r})}), v_{3,4}=(\pm1,0), v_{5,6}=(\pm\sqrt{1+(\overline{r}/C)^{1/(n-1)}},0)\}$ while for $n$ odd,
there are four degenerate minima, ${\cal M}_o =\{v_{1,2},v_{3,4}\}$. For all models in the above family, we can find the static solutions
\be\label{phi11}
\phi(x)=\pm\tanh(\overline{r} x)\hspace{0.1cm},\hspace{0.1cm}
\chi(x)=\pm\sqrt{\frac{2C}{n\overline{r}}}\,{\rm sech}^n(\overline{r} x )\,,
\ee
which connect the minima $v_{3}$ and $v_{4}$ by means of the orbit \eqref{orbita} given by the algebraical curve \be
\overline{r} n\chi^2=2C(1-\phi^2)^n\, ,
\ee
see Figure 4. These solutions carry the energy
\be
E_{\rm BPS}=4 \overline{r}/3+ \sqrt{\pi}C \Gamma[n+1]/\Gamma[n+\frac{3}{2}] \, .
\ee

\begin{figure}[h]
\includegraphics[height=3cm]{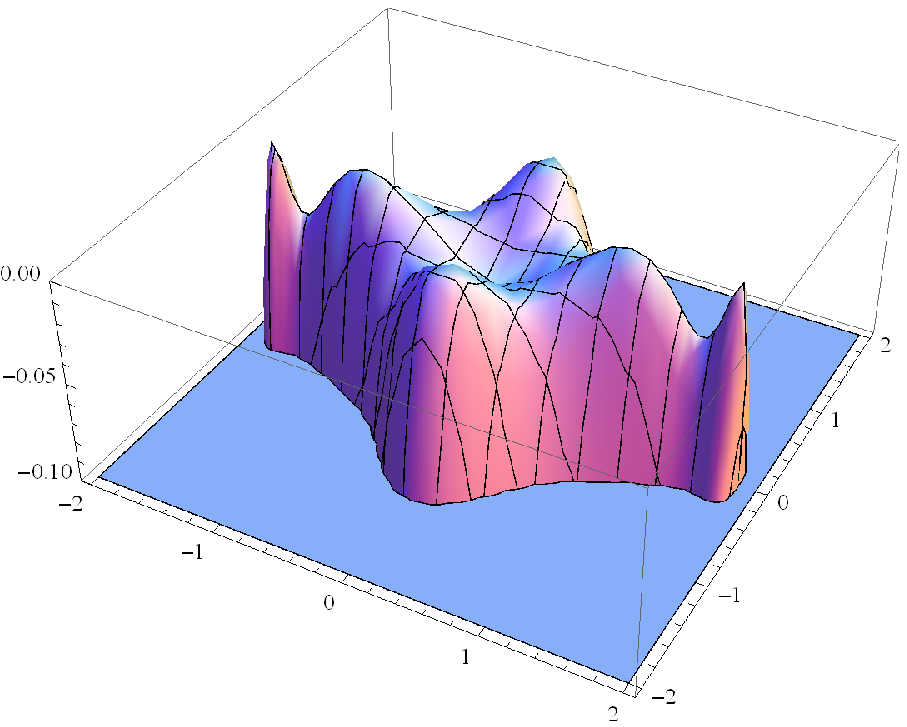} \hspace{0.5cm}
\includegraphics[height=3cm]{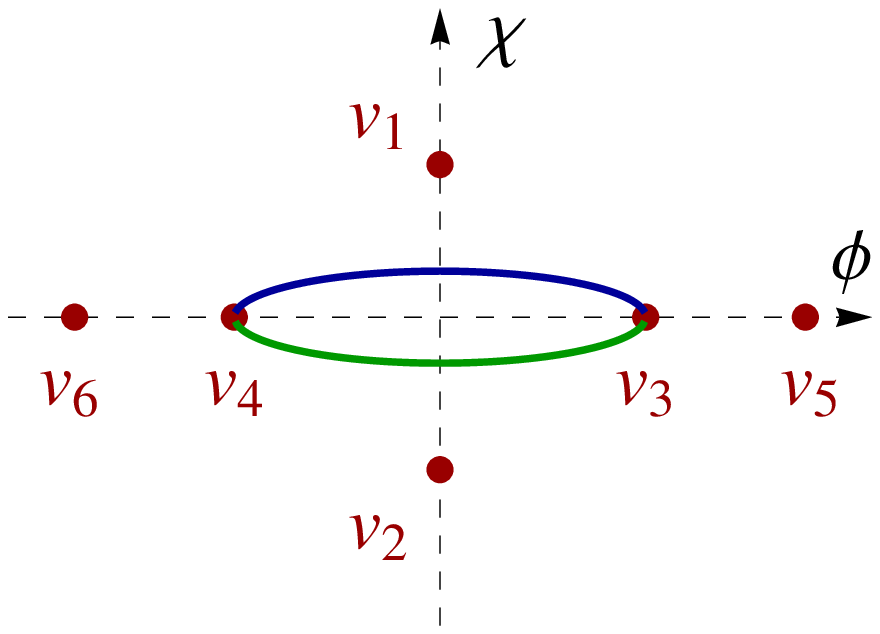}
\caption{Graphics of the potential $V(\phi,\chi)$ with $n=4$ and orbits of the kinks \eqref{phi11} in the internal plane.}
\end{figure}

Now, for the orbit $\chi(x)=0$,
in the BPS sectors connecting neighbor minima at the $\phi$-axis, the solutions are obtained from \eqref{orbita0}
with $f(\phi)$ given by \eqref{fp1}, which may be solved case by case.  For $n=1$, we have the solution $\phi(x)=\pm\tanh\left[(\overline{r}+C) x\right]$, and for $n=2$ we get the implicit expression
\be
\frac{(1+\phi)}{(1-\phi)}\left(\frac{\sqrt{1+\frac{\overline{r}}{C}}-\phi}{\sqrt{1+\frac{\overline{r}}{C}}+\phi} \right)^{\sqrt{\frac{C}{C+\overline{r}}}}=e^{2\overline{r}x}
\ee
and so on for other values of $n$. It is remarkable that we can obtain the explicit expression for the two-component kinks \eqref{phi11} for any value of $n$ but not for the one-component kinks.

\subsubsection{Another family of models}

Here we take $b=0$, $\kappa=4r/n$, $a>0$, and integer $n>0$ and we redefine the coupling constant $r$ as $r=(\overline{r} n)/4$. From \eqref{fp1} we have
\be\label{fp11}
f(\phi)=(a-\phi)\left(\overline{r} \phi+C(a-\phi)^{n-1}\right)\,,
\ee
which generate a family of models whose potentials have up to $n+2$ minima: two minima at  $v_{1,2}=\left(0,\pm 2\sqrt{(a^n C)/(\overline{r} n)}\right)$,  one at $v_{3}=(a,0)$, and up to $n-1$ minima (for $n>1$) coming from the condition $g(\phi)=\overline{r} \phi+C(a-\phi)^{n-1}=0$.

For all models of this family, we get from \eqref{orbita}, \eqref {phi1} and \eqref{u11} in the sector connecting the minima $v_{1,2}$ and $v_{3}$, the static solutions
\be\label{phi12}
\phi(x)=\frac{a}{2}\left(1\pm \tanh\left(\frac{a\overline{r} x}{2}\right)\right)\,,
\ee
and
\be\label{chi12}
\chi(x)=\pm2^{1-\frac{n}{2}}\sqrt{\frac{c}{\overline{r} n}}\,a^\frac{n}{2} \, \left(1-\tanh\left(\frac{a\overline{r} x}{2}\right)\right)^{{n}/{2}}\,,
\ee
whose energy is given by $E_{\rm BPS}=\frac{\overline{r}a^3}{6}+\frac{Ca^{n+1}}{n+1}$, see Figure 5.

\begin{figure}[h]
\includegraphics[height=3cm]{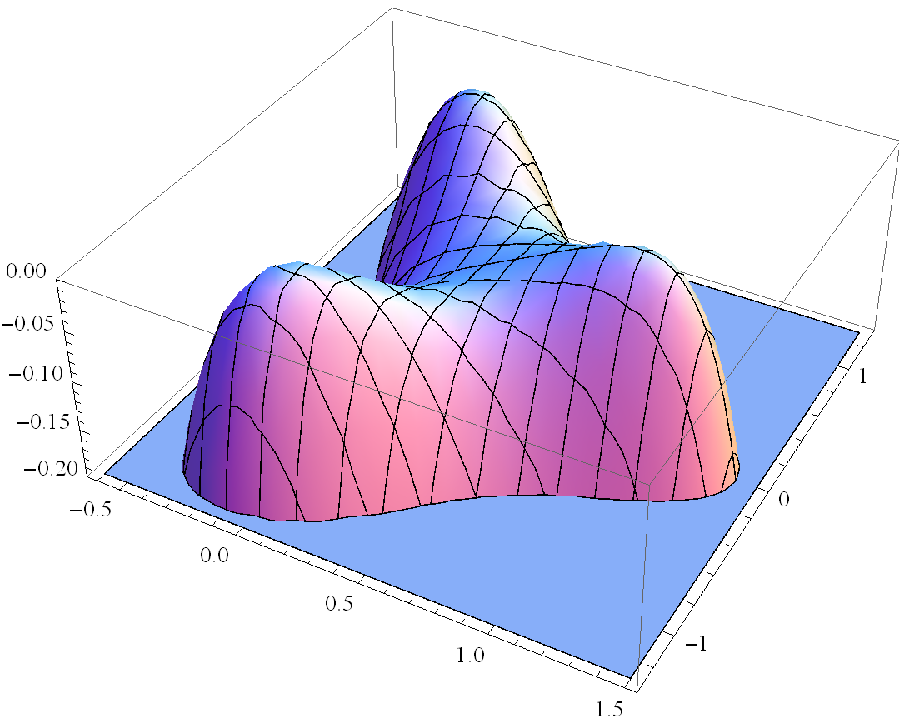} \hspace{0.5cm}
\includegraphics[height=3cm]{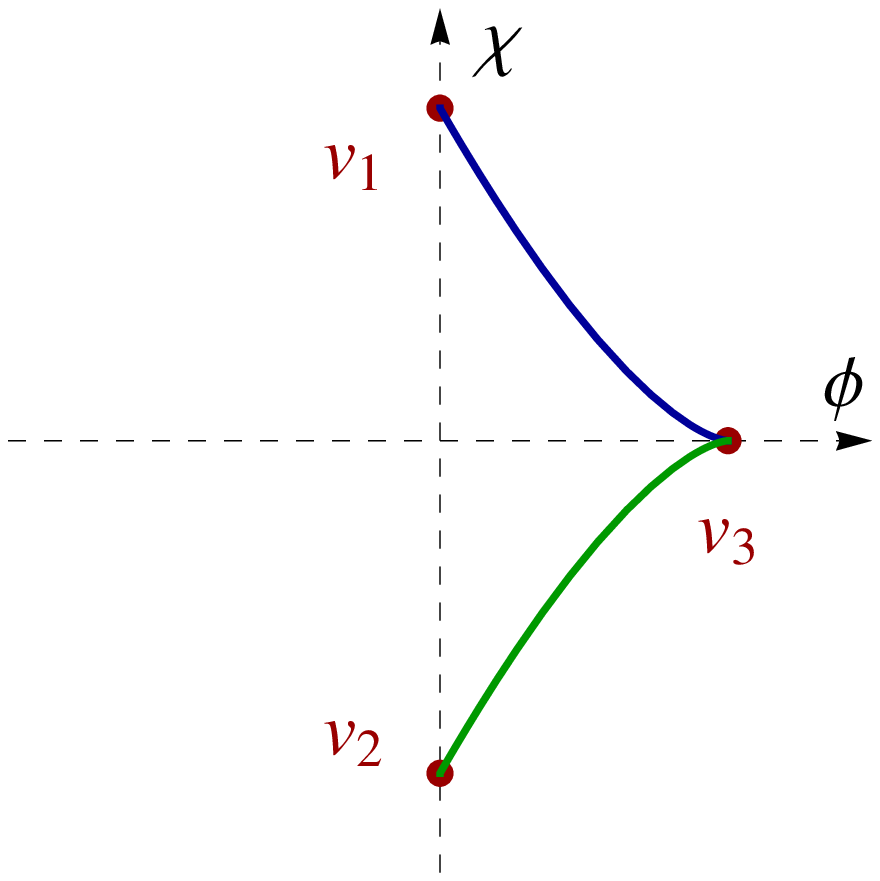}
\caption{Graphics of the potential $V(\phi,\chi)$ with $n=4$ and orbits of the kinks \eqref{phi12} and \eqref{chi12} in the internal plane.}
\end{figure}

\section{Nonpolynomial models}
\label{sec:4}

Let us now move on to the case of nonpolynomial potentials. Here we consider
\be
W(\phi,\chi)=\int^\phi{f(y)dy}-r\,\phi\,\sin\chi\, ,
\ee
such that the field potential is
\be \label{pot59}
V=\frac{1}{2} \left[ (f(\phi)-r\sin\chi)^2 + r^2 \phi^2 \cos^2\chi \right]\, ,
\ee
which is a periodic function in the variable $\chi$ as illustrated in Figure 6. The set of zeroes is given by ${\cal M}=\{v_{1,m_1},v_{2,m_2},v_{i+2,m_3} \}$ where
\ben
v_{1,m_1}&=&\Big(0, \arcsin \frac{f(0)}{r}+2\pi m_2\Big) \, , \nonumber\\
v_{2,m_2}&=&\Big(0, -\arcsin \frac{f(0)}{r}+\pi(2 m_3+1) \Big) \, , \nonumber\\
v_{i+2,m_3} &=& \Big(\phi^{(i)},\pi(m_1+\frac{1}{2})\Big) \, , \nonumber
\een
where $\phi^{(i)}$ are the roots of the function $f(\phi)-(-1)^n r$, and $m_i\in \mathbb{Z}$. In this case, the static solutions are obtained from the first-order equations
\be\label{gwp1s}
\phi^{\prime}=f(\phi)-r\,\sin\chi\hspace{0.3cm},\hspace{0.3cm}
\chi^{\prime}=-r\,\phi\,\cos\chi\,.
\ee
We can manipulate these equations to get
\be\label{ddp1s}
\phi^{\prime\prime}+\phi\,\phi^{\prime\,2}-\left(2\phi\,f(\phi)+\frac{df}{d\phi}\right)\phi^{\prime}-\phi\left(r^2-f^2(\phi)\right)\,=0\,.
\ee
Again, choosing $\phi(x)$ satisfying \eqref{u1u2}
we can rewrite the above Eq.\eqref{ddp1s} as
\be\label{uddps}
\frac{df}{d\phi}+P(\phi)f(\phi)-R(\phi)f^2(\phi)-Q(\phi)=0\,,
\ee
where
\be\label{pq1s}
P(\phi)=2\phi,\,\,\,\,R(\phi)=\frac{\phi}{U},\,\,\,\,Q(\phi)=\frac{dU}{d\phi}+\phi U-\frac{r^2 \phi}{U} \,.
\ee
The general solution is given by
\be\label{f76}
f(\phi)=U(\phi) + r\tanh \Big[ r \Big( C -\int \frac{\phi d\phi}{U(\phi)} \Big) \Big]
\ee
Then, choosing $U(\phi)$ we have
the constraint
\be\label{orbitas}
r\,\sin\chi=f(\phi)-U(\phi)\,,
\ee
for the field $\chi$.

Also, the first-order equations \eqref{gwp1s} supports the orbit
$\chi(x)=\pm(2n-1)\pi/2$, $n=0, 1, 2,...$. In this case, the static solutions $\phi(x)$, connecting neighbor minima located at the $\phi$-axis, are obtained from the equation
\be\label{orbita00}
\frac{d\phi}{dx}=f(\phi)+\,(-1)^n\,r\,.
\ee
Let us now illustrate the above procedure with an example. We start using
\be\label{u11s}
U(\phi)=\kappa(a-\phi)(b+\phi)\,,
\ee
where $a,b,\kappa$ are real parameters. By employing \eqref{f76} we obtain
\be\label{fp1s}
f(\phi)=U(\phi)+ r \frac{C^2(a-\phi)^{\frac{2ar}{\kappa(a+b)}} (b+\phi)^{\frac{2br}{\kappa(a+b)}}-1 }{C^2(a-\phi)^{\frac{2ar}{\kappa(a+b)}} (b+\phi)^{\frac{2br}{\kappa(a+b)}}+1}
\ee
Again we restrict ourselves to cases where the exponents in the previous expression are even integers, such that we impose that $\frac{ar}{\kappa(a+b)}=n$ with $n\in \mathbb{N}$, for instance by choosing $\kappa=\frac{ar}{n(a+b)}$. In this case
\be \label{f79}
f(\phi)=U(\phi)+ r \frac{C^2(a-\phi)^{2n} (b+\phi)^{2n}-1 }{C^2(a-\phi)^{2n} (b+\phi)^{2n}+1}
\ee
The field potential term is obtained by plugging \eqref{f79} into \eqref{pot59}. In spite of the complexity of this expression the kink solution is written by the expression \eqref{phi1} and from \eqref{orbitas}
\be
\chi = \arcsin \frac{C^2(a-\phi(x))^{2n} (b+\phi(x))^{2n}-1 }{C^2(a-\phi(x))^{2n} (b+\phi(x))^{2n}+1}
\ee
For example for $a=b=1$ we get
\ben
\phi&=& \tanh \frac{rx}{2n} \, , \label{kink88} \\
\chi&=& \arcsin \left(1- \frac{2}{1+C^2\,{\rm sech}^{2n}\,\frac{rx}{2n}} \right) \, ,\label{kink89}
\een
whose orbit is displayed in the Figure 6 for the case $n=2$.

\begin{figure}[h]
\includegraphics[height=3cm]{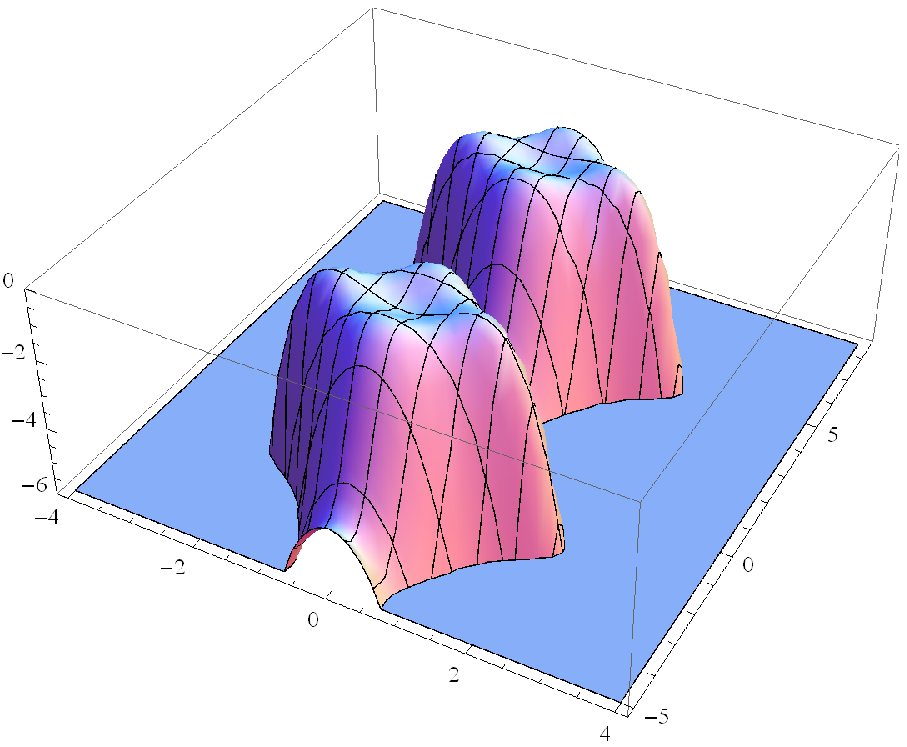} \hspace{0.5cm}
\includegraphics[height=3cm]{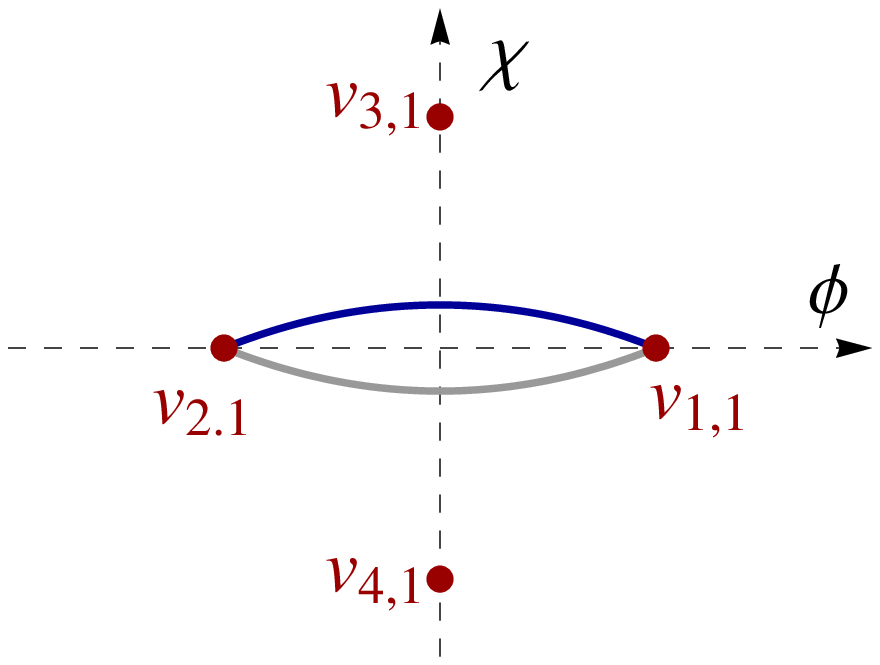}
\caption{Graphics of the potential $V(\phi,\chi)$ and orbits of the kinks \eqref{kink88} and \eqref{kink89} in the internal plane for $n=2$.}
\end{figure}

\section{Final Comments}
\label{sec:5}

In this work we proposed a new method to construct and solve models described by two real scalar field. The procedure is simple, inspired on the approach introduced in
Ref.~\cite{bd}, and it works for the construction of polynomial and non-polynomial models.

To illustrate the procedure, we studied several examples, which show how efficient the method is, in order to construct new two-field models with non-trivial two-component kink solutions. We note that the method starts with
$W$, the superpotential, so all the models we construct lead to first-order differential equations, which solve the equations of motion. In this sense, all the solutions we found are BPS states, and they are classically or linearly stable, as proved before in Ref.~\cite{bs}.

A relevant feature of the procedure is that it is different from the deformation procedure involving two-field models, and it is very simple to be applied in investigations based on two real scalar fields. An issue which deserves further examination concerns the extension of the method to three or more real scalar fields. This is under investigation, and we hope to report the new results in a separate work.

The authors would like to thank CAPES, CNPq and FAPESP, for partial financial support.


\end{document}